\documentclass[reqno]{amsart}

\theoremstyle{plain}
\newtheorem{theorem}{Theorem}[section]

\newtheorem{lemma}[theorem]{Lemma}
\newtheorem{corollary}[theorem]{Corollary}

\newtheorem*{quotethm}{}
\newtheorem*{notation}{Notation}
\newtheorem*{remark2}{Remark}

\DeclareMathOperator{\supp}{supp}
\DeclareMathOperator{\tr}{tr}

\numberwithin{equation}{section}

\newenvironment{acknowledgement}{\emph{Acknowledgement.}}

\newcommand\R{\mathbb R}
\newcommand\N{\mathbb N}
\newcommand\C{\mathbb C}
\newcommand\Z{\mathbb Z}
 
\newcommand\D{\mathcal{D}} 
\newcommand\Db{\mathbf{D}}

\newcommand\I{\mathcal{I}}

\newcommand\di{\mathrm d}

\newcommand{\Ec}{\mathcal{E}}

\newcommand\e{\mathrm{e}}
\newcommand\eps{\varepsilon}
\newcommand\vphi{\varphi}

\newcommand{\la}{\langle}
\newcommand{\ra}{\rangle}

\newcommand{\lt}{\left}
\newcommand{\rt}{\right}

\renewcommand\P{\mathbb P}
\newcommand\E{\mathbb E}

\newcommand{\abs}[1]{\left\lvert #1 \right\rvert}
\newcommand{\norm}[1]{\left\lVert #1 \right\rVert}

{\end{list}}

\begin{document}

\title[Delocalization in random Landau Hamiltonians]
{Dynamical delocalization in random Landau Hamiltonians}

\author{Fran\c cois Germinet}
\address{ Universit\'e de Cergy-Pontoise,
D\'epartement de Math\'ematiques,
Site de Saint-Martin,
2 avenue Adolphe Chauvin,
95302 Cergy-Pontoise cedex, France}
 \email{germinet@math.u-cergy.fr}

\author{Abel Klein}
\address{University of California, Irvine,
Department of Mathematics,
Irvine, CA 92697-3875,  USA}
 \email{aklein@uci.edu}

\author{Jeffrey H. Schenker}
\address{ETH Z\"urich,Theoretische Physik,
CH-8093 Z\"urich,
Switzerland}
 \email{jschenker@itp.phys.ethz.ch} 


\thanks{A.K. was supported in part   by NSF Grant
DMS-0200710}

\begin{abstract}
We prove the existence of dynamical delocalization for random Landau Hamiltonians
near each Landau level. 
Since typically there is dynamical localization at the edges of
each disordered-broadened Landau band, this implies the existence of at least
 one dynamical mobility edge at
each Landau band, namely a boundary  point between the localization and
delocalization regimes, which we prove to converge to the corresponding Landau
level as either the magnetic field or the disorder goes to zero. 
\end{abstract}

\maketitle

\section{Introduction}
\label{sectintro}

In this article we prove the existence of dynamical delocalization for 
random Landau Hamiltonians
near each Landau level.  More precisely, we prove that for these two-dimensional
 Hamiltonians there exists  at least one energy
$E$ near each Landau level such that $\beta(E)\ge \frac 14$,
 where  $\beta(E)$, the 
 local transport exponent introduced in \cite{GK5}, is a measure of the  rate of
transport for  which $E$ is responsible. 
Since typically there is dynamical localization at the edges of
each disordered-broadened Landau band, this implies the existence of at least
 one dynamical mobility edge at
each Landau band, namely a boundary  point between the localization and
delocalization regimes, which we prove to converge to the corresponding Landau
level as either the magnetic field or the disorder goes to zero.

Random Landau Hamiltonians are
the subject of intensive study  due to their  links with the
quantum Hall effect \cite{vKli}, for which von Klitzing received the 1985 Nobel
Prize in Physics.  They
 describe an  electron moving in  a very thin flat conductor
with impurities under the influence of a constant
 magnetic field perpendicular to the plane of the conductor, and 
 play an important role in the understanding of the quantum Hall effect
\cite{L,AA,T,Ha,NT,Ku,Be,ASS,BES}.  Laughlin's argument \cite{L}, as pointed out by
Halperin
\cite{Ha}, uses
 the assumption that under weak disorder and strong magnetic field the energy
spectrum consists of
 bands of extended states separated by energy regions of localized states and/or energy
gaps. (The experimental existence of a nonzero quantized Hall conductance was            
construed as evidence for the existence of extended states, e.g.,  \cite{AA,T}.)
 Halperin's analysis provided a theoretical justification
 for the existence of extended states
near  the Landau levels,
 or at least of some form of delocalization, and of nonzero Hall conductance. 
 Kunz \cite{Ku} stated assumptions  under
which  he derived the divergence of a  ``localization length" 
near each Landau level at weak disorder, in agreement with 
Halperin's argument. 
Bellissard, van Elst and Schulz-Baldes \cite{BES} proved that, for a
 random Landau Hamiltonian in  a tight-binding approximation,
if the Hall
conductance jumps from one integer value to another between two Fermi
energies, then there is an energy between these Fermi energies at which a
certain
localization length diverges.
Aizenman and Graf \cite{AG} gave a more elementary derivation of this result,
 incorporating  ideas of Avron,  Seiler and Simon \cite{ASS}.
  (We refer to \cite{BES} for an
 excellent overview of the quantum Hall effect.) But before the present paper
there were no results about non trivial transport 
and existence of a \emph{dynamical} mobility edge near 
the Landau levels.

The main open problem in random Schr\"odinger operators is delocalization,
the existence of ``extended states",  a forty-year old problem that
goes back to Anderson's seminal article \cite{An}.
 In three or more dimensions it is believed that there exists
 a transition  from an 
{insulator regime}, characterized by ``localized states", to a very
 different {metallic  regime} 
characterized by  ``extended states".
The energy at which this {metal-insulator transition} occurs is 
called the {``mobility edge"}. For two-dimensional  random
 Landau Hamiltonians  such a
 transition is expected
 to occur near each Landau level \cite{L,Ha,T}.

The occurrance of localization is by now well established, e.g., 
\cite{GMP,FS,FMSS,CKM,Sp,VDK,KLS,AM,FK1,A,Kl,CH1,CH2,FK,FK3,Wa,GDB,KSS,
CHT,ASFH,DS,
GK1,St,Wa2,Kl2,DSS,KK2,GK3,AENSS,BK} and many more. 
But delocalization is another story.  At present, the only mathematical result
 for a typical random Schr\"odinger operator 
(that is, ergodic and with a locally   H\"older-continuous  integrated density
of states at all energies)   is for
the Anderson model on the Bethe lattice, where Klein has proved 
that for small disorder the random operator has purely absolutely
continuous spectrum in a nontrivial  interval  \cite{K1} and exhibits 
ballistic behavior \cite{K2}.
   For
   lattice Schr\"odinger operators
with slowly decaying random potential, Bourgain proved existence of 
absolutely continuous 
spectrum in $d=2$ and constructed  proper extended
 states for dimensions $d\ge 5$ \cite{Bo1,Bo2}.
For lattice Schrodinger operators, Jaksic and Last \cite{JL} gave 
conditions under which the existence of 
  singular  spectrum can be ruled out, 
yielding the existence of absolutely continmuous spectrum.
Two other promising  approaches to the phenomena of delocalization 
 do not work directly with random Schr\"odinger operators;
one studies delocalization
in the context of random matrices
\cite{DPS,BMR,SZ}, and  the other studies a scaling limit of a random 
Schr\"odinger equation up to a disorder dependent finite time scale 
\cite{EY,Ch,ESY}.

But what do we mean by delocalization?   In the physics literature
one finds the expression
``extended states," which is often interpreted in the mathematics literature as
absolutely continuous spectrum.  But the latter may not be the correct interpretation
in the case of random Landau Hamiltonians;
 Thouless \cite{T}
discussed the possibility of singular continuous spectrum or even of the delocalization
occurring at a single energy.  In this paper we rely
on the approach to the metal-insulator transition developed by 
Germinet and Klein \cite{GK5},  based on transport intead of spectral
 properties.  It provides a structural 
result on the
dynamics of Anderson-type random operators:
 At a given energy $E$ there is either dynamical localization
($\beta(E)=0$)
 or  dynamical delocalization with a non zero minimal rate of tranport
($\beta(E) \ge \frac 1 {2d}$, with $d$ the dimension). An energy 
at which such a transition occurs is 
called a dynamical {mobility edge}. (The terminology used in this paper 
differs from the one in \cite{GK4,GK5}, which use strong insulator region for
the intersection of the region of dynamical localization with the spectrum, 
weak metallic region for the region of dynamical delocalization, and
 transport mobility edge for dynamical {mobility edge}.)

We prove that  for disorder and  magnetic
field for which the energy spectrum consists of  disjoint bands around the 
Landau levels (as in the case of  either  weak disorder or strong magnetic
field), the random Landau Hamiltonian   exhibits dynamical delocalization in
each band (Theorem~\ref{thmdeloc}).
 Since the existence of dynamical localization at the edges of the bands is
known \cite{CH2,Wa,GK3}, this proves the existence of dynamical mobility
edges.   We thus provide a mathematically rigorous derivation of 
 the previoulsly mentioned underlying assumption 
in  Laughlin's argument.

We also address the issue of the location of the delocalized energies in each 
disorder-broadened Landau band. Percolation arguments and numerical
results indicate that for large magnetic field there should be only one delocalized 
energy, located at the Landau level \cite{CD}.  We prove that these predictions
hold asymptotically.
That is, for the random Landau
 Hamiltonian studied in \cite{CH2,GK3}, we prove that  delocalized energies
 converge 
to the corresponding Landau level as the magnetic field goes to infinity
(Corollary~\ref{corlimit1}).
We also prove this result  as  the disorder goes to zero for 
an appropriately defined random Landau
 Hamiltonian (Corollary~\ref{corlimit2}).

Our proof of dynamical delocalization for random Landau Hamiltonians
is based on the use of some decidedly
nontrivial consequences  of the multiscale analysis for
 random Schr\"odinger operators combined with the generalized eigenfunction
 expansion  to establish properties of the Hall conductance.   It  
 relies on three main ingredients:

(1) The analysis in \cite{GK5} showing that for an Anderson-type random
 Schr\"odinger
operator   the region of dynamical localization
 is exactly the region of applicability
of the multiscale analysis, that is, the conclusions of the multiscale 
analysis are valid at every energy in the region of dynamical localization,
 and that outside this region some nontrivial transport must occur
with non zero minimal rate of transport.

(2)  The random Landau Hamiltonian 
  satisfies all the requirements for the multiscale analysis (i.e., the hypotheses
in \cite{GK1,GK5})  at all energies.  The only difficulty here is a
 Wegner estimate
 at all energies, including the Landau levels, 
a required hypothesis for
applying (1). If the single
bump in the Anderson-style potential  covers the unit square
 this estimate was known \cite{HLMW}. But if the single bump  has small support
  (which is the most interesting case for this paper in view of Corollary 2.3), a Wegner estimate 
 at all   energies was only known for the case of  rational flux in the unit square
\cite{CHK}.   We prove a new Wegner estimate which has no restrictions
on the magnetic flux in unit square (Theorem~\ref{thmWegner}).
This Wegner estimate holds in appropriate squares with integral
flux, hence the length scales of the squares may not be commensurate
with  the distances between bumps in the Anderson-style potential. 
This problem is overcome by performing the multiscale analysis
with   finite volume operators defined with boundary conditions
depending on the location of the square (see the discussion in 
Section~\ref{sectWegner}).

(3) Some information on the Hall conductance, namely:  (i)  The precise values
of the Hall conductance for the (free) Landau Hamiltonian: it is constant
between Landau levels and jumps by one at each Landau level,   a well
known fact (e.g., \cite{ASS,BES}).
(ii) The Hall conductance is constant as a function of the disorder
parameter in the gaps between the spectral bands around the Landau levels,
a result derived by Elgart and Schlein \cite{ES}  for smooth potentials
and extended here to more general potentials 
(Lemma~\ref{lemconducgap}).
 Combining  (i) and (ii)
we  conclude that the Hall conductivity  cannot be constant 
across Landau levels.
(iii) The Hall conductance is well defined and  
constant in intervals of  dynamical localization.
This is proved here  in a very transparent way
using a deep consequence of the multiscale analysis, called
 SUDEC,  derived from a new characterization of the region of dynamical 
localization \cite{GK6}.
 SUDEC  is used to 
show that
 in intervals of dynamical
localization the change in the Hall conductance is given by the (infinite) sum 
of the contributions to the Hall conductance due to the individual localized
 states, which is trivially seen to be equal to zero.
(See Lemma~\ref{lemconduc}.  This constancy in intervals of localization was
known for discrete operators as a consequence of the quantization of the Hall
conductance \cite{BES,AG}.  An independent but somewhat similar  proof
for discrete operators with finitely degenerate 
eigenvalues is found in the recent paper
 \cite{EGS}.   We do not need to rule out eigenvalues  with infinite multiplicity
for random Landau Hamiltonians; they are controlled using SUDEC.)
  Combining with (i) and (ii), we will conclude that there must be
dynamical delocalization as we cross a Landau level.

It is worth noting that each of the three ingredients (1), (2) and (3) is based
on  intensive
research conducted over  the past 20 years. (1)
 relies on the ideas of the multiscale analysis, originally
introduced by  Fr\"ohlich and  Spencer \cite{FS} and further developped in
\cite{FMSS,vD,VDK,Sp,CH1,FK,GK1}.   
(2), namely the Wegner estimate, originally proved for lattice operators by 
 Wegner \cite{W}, is a key tool for the multiscale analysis, and it has been 
 studied in  the continuum in \cite{CH1,Kl,HLMW,CHN,HK,CHK}.  (3) 
has  a long story in
the study of the quantum Hall effect
 \cite{L,Ha,TKNN,Ku,Be,ASS,BES,AG,ES,EGS}.
 
In this paper we give a  simple and self-contained 
analysis of the Hall conductance based on  consequences of
localization for random Schr\"odinger operators.  In particular, we do not
use the fact that the quantization of the Hall conductance 
is a consequence of the geometric interpretation of this quantity
as a Chern character or a Fredholm index  [TKNN, Be, AvSS, BES, AG].
Our analysis applies
when the disorder-broadened Landau levels do not overlap
 (true at either large magnetic field
 or small disorder).  
In a sequel, extending an argument of [AG] for discrete
operators, we will discuss quantization of the Hall conductance for ergodic
Landau Hamiltonians in the region where we have sufficient decay of operator
kernels of the Fermi projections.  
 This fact is well known for lattice
Hamiltonians [Be, BES, AG], but the details of the proof have been spelled out
for continuum operators only in spectral gaps [AvSS].  Combining results from
 the present paper and its sequel we
expect to prove dynamical delocalization for random  Landau Hamiltonians
 in cases  when the Landau bands overlap.

This paper is organized as follows:  In Section~\ref{sectresult}
we introduce the random Landau Hamiltonians and state our  results.
Our main result is Theorem~\ref{thmdeloc},
the existence of dynamical delocalization for  random Landau Hamiltonians
near each Landau level. This theorem is restated in a more general form as
 Theorem~\ref{corbeta}, which is proved in Section~\ref{sectconduc}.   In Corollary~\ref{corlimit1} we give a rather complete
 picture for 
 random Landau Hamiltonians at large magnetic field  as in \cite{CH1,GK3}:
there are dynamical mobility edges in each Landau level, which
converge to the
corresponding Landau level as the magnetic field goes to infinity.
Corollary~\ref{corlimit2} gives a similar picture as the disorder goes to zero;
it is proven in Section~\ref{sectproof}.
In Section~\ref{sectWegner} we show that random Landau Hamiltonians
satisfy all the requirements for a multiscale analysis; 
Theorem~\ref{thmWegner} is the Wegner estimate.

\begin{notation} We write $ \langle x \rangle := \sqrt{1+|x|^2} $. The characteristic function of a set $A$ will be denoted by $\chi_A$.
Given $x \in \R^2$ and $L >0$ we set
 $$ \Lambda_L(x) :=\left \{y\in\mathbb{R}^2  ; \ \,
 |y-x|_\infty<{\tfrac L 2}\right\}, \quad 
 \ \chi_{x,L} := \chi_{\Lambda_L(x)}, \quad \chi_x:= \chi_{x,1}.  $$
$C^\infty_c (I)$ denotes  the class
of  real valued infinitely differentiable functions on 
$\mathbb{R}$ with compact
support contained in the open interval  $I$, with $C^\infty_{c,+} (I)$ being
 the subclass of nonnegative functions.
The Hilbert-Schmidt norm of an operator $A$ is written as
$\|A\|_2 =  \sqrt{\tr  A^*A}$. 
\end{notation}

\begin{acknowledgement}  The authors are grateful to  Jean Bellissard,
Jean-Michel Combes, 
Peter Hislop and Fr\'ederic Klopp for many helpful  discussions.
\end{acknowledgement}

\section{Model and results}
\label{sectresult}

We consider the random Landau Hamiltonian 
\begin{equation} \label{landauh}  H_{B,\lambda,\omega} = H_B +
\lambda  V_\omega \quad \mathrm{on} \quad
\mathrm{L}^2(\mathbb{R}^2, {\mathrm{d}}x), 
\end{equation}
where $H_B$ is the (free) Landau Hamiltonian,
 \begin{equation}
H_B =  (-i\nabla-\mathbf{A})^2 \quad \text{with} \quad 
\mathbf{A}= \tfrac B2 (x_2,-x_1).
\end{equation} 
Here $\mathbf{A}$ is the vector potential and
$B>0$ is the strength of the magnetic field, we use the symmetric gauge
and incorporated the charge of the electron in the vector potential).
The parameter
$\lambda > 0$ measures the disorder strength,  and
 $V_\omega$ is a  random potential of the form
\begin{equation} \label{potVL}
 V_\omega(x) = \sum_{i \in\mathbb{Z}^2} \omega_i\, u(x-i)  ,
\end{equation} 
with  $u$ a measurable function satisfying
$ u^{-} \chi_{0,\eps_u}\le u \le u^+ \chi_{0,\delta_u}$ for some 
$0<\eps_u\le \delta_u<\infty$ and $0< u^{-}\le  u^{+}<\infty$,
and $\omega =\{\omega_i; \ i\in\mathbb{Z}^2\}$ a 
family of independent,
identically distributed random variables taking values in a bounded interval 
$[-M_1, M_2]$ ($0 \le M_1,M_2 < \infty$, $ M_1 + M_2 > 0$) whose common probability
distribution $\nu$ has a bounded density $\rho$.  (We write $(\Omega,\P)$ for the underlying
probability space, and
$\E$ for the corresponding
expectation.)  Without
 loss of generality
 we set $\left\|\sum_{i \in\mathbb{Z}^2} \, u(x-i)\right\|_\infty=1$, and hence
  $\, -M_1 \le   V_\omega(x)\le M_2$.

$ H_{B,\lambda,\omega}$ is a random 
operator, i.e., the mappings $\omega \to f( H_{B,\lambda,\omega})$ are strongly
measurable for all bounded measurable functions on $\mathbb{R}$.  We define
the magnetic translations $U_a= U_a(B)$,  $ a \in \R^2$, by
\begin{equation}\label{magtrans}
\left(U_a \psi\right)(x) = \e^{-i \frac B2  (x_2a_1 - x_1 a_2)} \psi(x -a),
\end{equation}
obtaining a projective unitary representation of 
$\R^2$ on $\mathrm{L}^2(\mathbb{R}^2, {\mathrm{d}}x)$:
\begin{equation}
U_a U_b =   \e^{i \frac B2  (a_2b_1 - a_1 b_2)}  U_{a+b} =
\e^{i  B  (a_2b_1 - a_1 b_2)} U_b U_a, \quad a,b \in \Z^2. 
\end{equation}
We have  $U_a H_{B} U_a^* = H_B$ for all $a \in \R^2$, 
and for magnetic translation by elements of $\Z^2$
we have the covariance relation:
\begin{equation}\label{covariance}
U_a H_{B,\lambda,\omega} U_a^* = H_{B,\lambda,\tau_a\omega}
 \quad \text{for $a \in \Z^2$},
\end{equation}
where $ (\tau_a\omega )_i = \omega_{i -a} $, $i \in \Z^2$.  It follows that   
$H_{B,\lambda,\omega}$ is a  $\mathbb{Z}^2$-ergodic  random
self-adjoint operator on  
$\mathrm{L}^2(\mathbb{R}^2,{\rm d}x)$;
hence
 there exists a nonrandom set $\Sigma_{B,\lambda}$ such that
 $\sigma (H_{B,\lambda,\omega})=\Sigma_{B,\lambda} $ with probability one,
 and  the
decomposition of $\sigma (H_{B,\lambda,\omega})$ into pure point spectrum,
absolutely continuous spectrum, and singular continuous spectrum is also
independent of the choice of $\omega $ with probability one \cite{KM,PF}.

The spectrum $\sigma(H_B)$  of the  Landau
Hamiltonian $H_B$ consists of a sequence of infinitely
degenerate eigenvalues, the
Landau levels:
\begin{equation} \label{landaulevels}
 B_n=(2n-1)B ,\quad n=1,2,\dotsc .
\end{equation}
It will be convenient to set  $B_0=-\infty$. A simple argument shows  that 
\begin{equation} \label{splandau}
\Sigma_{B,\lambda} \subset
  \bigcup_{n=1}^\infty \mathcal{B}_n(B,\lambda) ,\quad\mbox{where}
\quad \mathcal{B}_n(B,\lambda)=
 [B_n - \lambda M_1, B_n +\lambda M_2]  . 
\end{equation}
If   the
 \emph{disjoint bands condition} 
\begin{equation} \label{gapcond}
\lambda {(M_1 + M_2)}<  {2B},
\end{equation}
is satisfied (true at either weak disorder or strong magnetic
field), the (disorder-broadened) Landau  bands $\mathcal{B}_n(B,\lambda)$ are
 disjoint, and hence the open intervals
\begin{equation}\label{Gn}
\mathcal{G}_n(B,\lambda)=]B_n +\lambda M_1, B_{n+1} - \lambda M_2[, \quad
n=0,1,2,\dotsc ,
\end{equation}
are nonempty spectral gaps for $H_{B,\lambda,\omega}$. Moreover,
 if $\rho > 0$ a.e.
on  $[-M_1, M_2]$ and \eqref{gapcond} holds, then  for each $B>0$, $\lambda >0$, and $n=1,2,\ldots$
 we can find 
$a_{j,B,\lambda,n} \in [0,\lambda M_j]$,
$j=1,2$, 
 continuous in $\lambda$, 
 such that \cite[Theorem 4]{KM2}
\begin{equation} \label{splandau2}
\Sigma_{B,\lambda} =
  \bigcup_{n=1}^\infty \mathcal{I}_n(B,\lambda), \
\quad \mathcal{I}_n(B,\lambda)=
 \left[B_n - a_{1,B,\lambda,n},B_n+ a_{2,B,\lambda,n}\right] . 
\end{equation}

Our   main result says that under the disjoint bands condition the random 
Landau Hamiltonian  
$H_{B,\lambda,\omega}$
exhibits dynamical delocalization in each Landau
 band  $\mathcal{B}_n(B,\lambda)$.
To measure ``dynamical delocalization" we introduce
\begin{equation}
 M_{B,\lambda,\omega}(p,\mathcal{X},t)  = 
 \left\|  {\langle} x {\rangle}^{\frac p 2}
 {\mathrm{e}^{-i tH_{B,\lambda,\omega} }}
\mathcal{X}(H_{B,\lambda,\omega}) \chi_0 
 \right\|_2^2 ,
\end{equation}
the
 random moment of order
$p\ge 0$ at time $t$ for the time evolution  in the Hilbert-Schmidt norm,  
initially spatially localized in the square of side one around the origin
(with characteristic function $\chi_0$), and ``localized" 
in energy by the function $\mathcal{X}\in C^\infty_{c,+} (\mathbb{R})$.
Its time averaged expectation is given by
\begin{equation}   \label{tam}
 \mathcal{M}_{B,\lambda}( p ,\mathcal{X}, T )   =  
\frac1{T} \int_0^{\infty}
 \mathbb{E}\left\{ M_{B,\lambda,\omega}(n,\mathcal{X},t)\right\}
{\mathrm{e}^{-\frac{t}{T}}} \,{\rm d}t .
\end{equation}

\begin{theorem}\label{thmdeloc}
Under the disjoint bands condition the random Landau Hamiltonian  
$H_{B,\lambda,\omega}$ 
exhibits   dynamical delocalization
  in each Landau band $\mathcal{B}_n(B,\lambda)$: 
For each  $n=1,2,\ldots$  there exists at least one energy 
 $E_n(B,\lambda)\in \mathcal{B}_n(B,\lambda) $,
such that for every  
 $\mathcal{X}\in C^\infty_{c,+} (\mathbb{R})$  with
 $\mathcal{X} \equiv 1$  on some open interval  $J\ni E_n(B,\lambda) $
 and  $p>0$,  we have 
\begin{equation}\label{momentgrowth}
\mathcal{M}_{B,\lambda}(p,\mathcal{X}, T)    \ge \
C_{p,\mathcal{X}} \, T^{\frac p4 - 6}
\end{equation}
for all  $T \ge 0$  with  $  C_{p,\mathcal{X}} > 0 $.
\end{theorem}

Following \cite{GK5}, we introduce the  (lower) transport  exponent 
\begin{eqnarray} 
 \label{betanX}
 \beta_{B,\lambda} (p,\mathcal{X}) =
\liminf_{T\to\infty} \; \frac{\log_+ {\mathcal{M}_{B,\lambda}}(p,\mathcal{X},T)}{  p 
\log T} \ \ \text{for $p\ge 0$, 
 $ \mathcal{X}\in C^\infty_{c,+} (\mathbb{R})$}, 
\end{eqnarray} 
where $\log_+ t = \max \{\log t,0\}$, and define the  \emph{$p$-th local 
transport exponent} at the energy $E$  by ($I$ denotes an open interval) 
\begin{equation}
\beta_{B,\lambda} (p,E)=  \inf_{I \ni E}
\sup_{\mathcal{X}\in C^\infty_{c,+} (I)} \beta_{B,\lambda} (p,\mathcal{X}).
\end{equation}
 The  transport exponents
$\beta_{B,\lambda} (p,E)$  provide
a measure of the rate of transport for which $E$ is responsible.  They 
are increasing in $p$ and hence 
we define the \emph{local (lower) transport exponent}
 $\beta_{B,\lambda} (E)$ by
\begin{equation} \label{defbetaE=intro}
\beta_ {B,\lambda} (E) = \lim_{p\to \infty} \beta_{B,\lambda}(p,E) = 
\sup_{p>0} \beta_{B,\lambda}(p,E)  .
\end{equation}
These transport exponents satisfy  the ballistic bound 
 \cite[Proposition 3.2]{GK5}: 
$ 0 \le \beta_{B,\lambda} (p,\mathcal{X}), \,
\beta_{B,\lambda} (p,E),\, \beta_{B,\lambda}(E) \le 1$.  Note that
$\beta_{B,\lambda}(E) =0$ if
$E \notin \Sigma_{B,\lambda}$.

Using this local transport exponent we define two complementary regions in the energy axis for fixed
 $B>0$ and $\lambda>0$:
the \emph{region of dynamical localization},
\begin{equation}\label{XiDL}
\Xi_{B,\lambda}^{{\text{DL}}}= \left\{E \in \R; \quad  \beta_{B,\lambda}(E)=0  \right\},
\end{equation}
and the region of \emph{dynamical delocalization},
\begin{equation}\label{XiDD}
\Xi_{B,\lambda}^{\text{DD}}= \left\{E \in \R; \quad  \beta_{B,\lambda}(E)>0  \right\}.
\end{equation}
Note that $\Xi_{B,\lambda}^{DL}$ is an open set and that 
$\Xi_{B,\lambda}^{\text{DD}} \subset \Sigma_{B,\lambda}$. 

We may now restate Theorem~\ref{thmdeloc} in a more general form as

\begin{theorem}\label{corbeta} Consider
a random Landau Hamiltonian  
$H_{B,\lambda,\omega}$ under the disjoint bands condition \eqref{gapcond}.
Then for all $n=1,2,\ldots$ we have  
\begin{equation}\label{notempty}
\Xi_{B,\lambda}^{\text{DD}} \cap \mathcal{B}_n(B,\lambda)\not=
 \emptyset.
\end{equation}
In particular,   there exists  at least one energy
 $E_n(B,\lambda)\in \mathcal{B}_n(B,\lambda) $ satisfying
\eqref{momentgrowth} and
\begin{equation}\label{betas}
 \text{${\beta_{B,\lambda}}(p,E_n(B,\lambda) )\ge \frac1{4} - \frac{11}{2p}>0 $ for all 
$p > 22$ and $\beta_{B,\lambda}(E_n(B,\lambda))\ge \frac14$}.
\end{equation}
\end{theorem}

Theorem~\ref{corbeta} 
is proved in Section~\ref{sectconduc}.
We will prove \eqref{notempty}, from which 
\eqref{betas} and \eqref{momentgrowth} 
 follows by  \cite[Theorems~2.10 and  2.11]{GK5}.

Next we investigate the location of the delocalized energy $E_n(B,\lambda)$,
and show in two different asymptotic regimes that it converges to the $n$-{th}
Landau level.   We recall that in the physics literature
localized and extended states are expected to be separated
by an energy called a mobility edge. 
Similarly, there is a natural definition for a \emph{dynamical mobility edge}:
 an energy $\tilde{E} \in \Xi_{B,\lambda}^{\text{DD}}
 \cap \left\{\overline{\Xi_{B,\lambda}^{DL} \cap 
\Sigma_{B,\lambda}}\right\}$, that is, 
an energy where the spectrum undergoes a transition from 
 dynamical localization to 
dynamical delocalization.

 In the regime of large magnetic field (and fixed disorder)
 we have the following rather complete picture for the model studied in 
\cite{CH2,GK3},  consistent with the prediction that at very large magnetic
field there is only one delocalized energy in each Landau band, located at the
Landau level \cite{CD}.

\begin{corollary}\label{corlimit1}Consider
a random Landau Hamiltonian  
$H_{B,\lambda,\omega}$  satisfying the following additional conditions
on the random potential:
 (i) $u \in C^2$ and $\supp u\subset D_{\frac {\sqrt{2}} 2}(0)$, the 
open  disc of radius $\frac {\sqrt{2}} 2$
centered at $0$.
(ii) The density of the probability distribution $\nu$ is an 
even function
$\rho >0$ a.e. on $[-M,M]$ ($M=M_1=M_2$).
(iii)
$\nu([0,s])\ge c \min\{s,M\}^\zeta$ for some $c>0$ and $\zeta>0$.
Fix $\lambda>0$ and let $B>0$ satisfy \eqref{gapcond}, in which case the spectrum
 $\Sigma_{B,\lambda}$ is given by \eqref{splandau2} with
\begin{equation}  \label{ab}
0 \le \lambda M - a_{j,B,\lambda,n}\le 
C_n(\lambda) B^{-\frac 12},\quad \text{$j=1,2$}.
\end{equation}
Then for
all $n=1,2,\dots$, if $B$ is large enough there exist dynamical mobility edges 
$\widetilde{E}_{j,n}(B,\lambda)$, 
$j=1,2$, with
\begin{gather}  \max_{j=1,2} \left \lvert  \widetilde{E}_{j,n}(B,\lambda)  - B_n  \right \rvert  
\le  K_n(\lambda)\frac{\log B}  B  \to  0 \quad \text{as $B \to \infty$},
\label{B2infty} \\
\label{loctildeE}
B_n - a_{1,B,\lambda,n}<\widetilde{E}_{1,n}(B,\lambda)\le 
\widetilde{E}_{2,n}(B,\lambda)<
B_n +a_{2,B,\lambda,n},\\
\label{abc}
 [B_n - a_{1,B,\lambda,n},\widetilde{E}_{1,n}(B,\lambda)[ \,\cup\,
]\widetilde{E}_{2,n}(B,\lambda), B_n + a_{2,B,\lambda,n}] \subset
\Xi_{B,\lambda}^{\text{DL}}.
\end{gather}
(By $C_n(\lambda), K_n(\lambda) $ we denote finite constants.  It is possible
 that  $\widetilde{E}_{1,n}(B,\lambda)=
\widetilde{E}_{2,n}(B,\lambda)$, i.e., dynamical delocalization occurs at 
a single energy.)
\end{corollary}

\begin{proof}  The estimate \eqref{ab} is proven in \cite{CH2},
the existence of energies  $\widetilde{E}_{j,n}(B,\lambda)$, $j=1,2$,
satisfying  \eqref{loctildeE}, \eqref{abc}
and \eqref{B2infty} is proven in \cite[Theorem~4.1]{GK3}. The fact that
we can choose
$\widetilde{E}_{j,n}(B,\lambda)$, $j=1,2$, that are also 
dynamical mobility edges follows from Theorem~\ref{thmdeloc}.
\end{proof}

We now investigate the small disorder regime (at fixed magnetic field) and prove
a result in the spirit of Corollary~\ref{corlimit1}. It is not too  interesting to
just let
 $\lambda \to 0$  in \eqref{landauh}, since the spectrum of the Hamiltonian
would then shrink
 to the Landau levels (see \eqref{splandau}) and the result would be trivial. In
order to keep the size of the spectrum constant we  rescale the probability
distribution $\nu$ of the $\omega_i's$ by concentrating more and more of
the mass of $\nu$ around zero as $\lambda \to 0$.

\begin{corollary}\label{corlimit2}  Let $\rho >0$ a.e on $\R$ be the density of
a probability distribution
$\nu$ with $\la u \ra^\gamma \rho(u)$ bounded for some $\gamma > 1$.
   Fix $b>0$, and set
$ \nu_\lambda $ to be the probability distribution with density
$\rho_\lambda (u) =  c_{b,\lambda} \lambda^{-1}\rho(\lambda^{-1}
u)\chi_{[-b,b]}(u)$, where the constant $c_{b,\lambda}$ is chosen so that 
$\nu_{\lambda}(\R)=\nu_{\lambda}([-b,b])=1$. Define
$H_{\omega,B,\lambda}$ by (2.1) with $\lambda =1$ but with the $\lambda$
dependent common probability distribution $\nu_\lambda$ for the random
variables $\{\omega_i; \, i \in \Z^2\}$. Assuming
$B>b$,
\eqref{gapcond} holds and the spectrum
 $\Sigma_{B,\lambda}$  given by  \eqref{splandau2}
 is independent of $\lambda$.
Then for
all $n=1,2,\dots$, if $\lambda$ is small enough there exist dynamical mobility edges 
$\widetilde{E}_{j,n}(B,\lambda)$, 
$j=1,2$, satisfying \eqref{loctildeE} and \eqref{abc},
and  we have 
\begin{align} 
 \max_{j=1,2} \left \lvert  \widetilde{E}_{j,n}(B,\lambda)  - B_n  \right \rvert  
\le  K_n(B)\lambda^{\frac {\gamma -1} \gamma}
 \abs{\log \lambda}^{\frac 2 \gamma} \to  0 \quad 
\text{as $\lambda \to 0$},
\label{lambda20}
\end{align}
 with a finite  constant  $ K_n(B) $. 
Moreover,  if  the density $\rho$ satisfies the stronger condition of  $\e^{\abs{u}^\alpha} \rho(u)$
being bounded for some  $\alpha>0$, the estimate in  \eqref{lambda20} holds with   
$  K_n(B)\lambda
 \abs{\log \lambda}^{\frac 1 \alpha}$ in the right hand side. 
  (It is possible
 that  $\widetilde{E}_{1,n}(B,\lambda)=
\widetilde{E}_{2,n}(B,\lambda)$, i.e., dynamical delocalization occurs at 
a single energy.)
\end{corollary}

Corollary~\ref{corlimit2} is proven in Section~\ref{sectproof}.

\section{The existence of dynamical delocalization}
\label{sectconduc}

In this section we prove Theorem~\ref{corbeta}
 (and hence  Theorem~\ref{thmdeloc}). 
For convenience we write   $H_{B,0,\omega}=H_B$ and extend \eqref{XiDL} to
$\lambda=0$ by 
$\Xi_{B,0}^{\text{DL}}=
\R \backslash \sigma(H_B)= \R \backslash \{B_n; \ n=1,2,\dots\}  $;
 the statements  below will hold (trivially) for $\lambda=0$ unless
this case is explicitly excluded.   Given a Borel set $\mathcal{J}\subset \R$,
 we set
$P_{B,\lambda,\mathcal{J},\omega} =
\chi_{\mathcal{J}}(H_{B,\lambda,\omega})$.  If
$\mathcal{J}= ]-\infty,E]$, 
we write   $P_{B,\lambda,E,\omega}$ for $P_{B,\lambda,]-\infty,E],\omega}$,
the Fermi projection corresponding to the Fermi energy $E$.

The random Landau Hamiltonian  $H_{B,\lambda,\omega}$ ($\lambda > 0$)
  satisfies all the hypotheses
in \cite{GK1,GK5}  at all energies, 
as shown in  Section~\ref{sectWegner}. 
The  following results are relevant to  the proof of  Theorem~\ref{corbeta}:
RDL (region of dynamical localization), 
RDD (region of dynamical delocalization), DFP (decay of the Fermi projection),
and SUDEC (summable uniform decay of eigenfunction correlations).

\begin{quotethm}[\textbf{RDL}]  The region of dynamical localization
$\Xi_{B,\lambda}^{\text{DL}}$ is exactly the region of applicability
of the multiscale analysis, that is, the conclusions of the multiscale 
analysis are valid at every energy $E \in \Xi_{B,\lambda}^{\text{DL}}$
 \cite[Theorem 2.8]{GK5}.
\end{quotethm}

\begin{quotethm}[\textbf{RDD}] Let  $\lambda > 0$.  If an energy $E$ is
in the region of dynamical delocalization $\Xi_{B,\lambda}^{\text{DD}}$
we must have ${\beta_{B,\lambda}}(E) \ge \frac 14$; in fact,
${\beta_{B,\lambda}}(p,E )\ge \frac1{4} - \frac{11}{2p}>0 $ for all 
$p > 22$.  Moreover, for each $\mathcal{X}\in C^\infty_{c,+} (\mathbb{R})$ with
$\mathcal{X} \equiv 1$ on some open interval $J\ni E $ we have
\begin{equation}\label{eqhyp}
\lim_{T \to \infty}\frac1{T^{\alpha}}
\mathcal{M}_{B,\lambda}(p,\mathcal{X},T)=\infty 
\end{equation}
for all $\alpha \ge 0$ and 
$ p>4\alpha+22$  \cite[Theorems 2.10 and 2.11]{GK5}.
\end{quotethm}

\begin{quotethm}[\textbf{DFP}]  The Fermi projection
 $P_{B,\lambda,E,\omega}$
exhibits fast decay if  the Fermi energy $E$ is in the region of
dynamical localization
$\Xi_{B,\lambda}^{\text{DL}}$: 
 If $E \in  \Xi_{B,\lambda}^{\text{DL}}$ and $\zeta \in ]0,1[$ we have
\begin{equation}\label{decayfermi}
\E \left\{\norm{\chi_x P_{B,\lambda,E,\omega} \chi_y}_2^2\right\} \le
 C_{\zeta,B, \lambda,E}\
 \e^{-|x-y|^\zeta} \quad \text{for all} \ x,y \in \Z^2 ,
\end{equation}
with the constant  $ C_{\zeta,B, \lambda,E}$ locally bounded in $E$.
\emph{(See \cite{GK6}, the result is based on \cite[Theorem 3.8]{GK1} and
\cite[Theorem 1.4]{BGK}.)}
As a consequence, 
 for $\P$-a.e.\ $\omega$ and  each  $\zeta \in ]0,1[$ there exists    
$C_{\zeta,B, \lambda,E,\omega}< \infty$, locally bounded in $E$, such that
\begin{equation}\label{decayFBC}
\norm{\chi_x P_{B,\lambda,E,\omega} \chi_y}_2 \le
 C_{\zeta,B, \lambda,E,\omega} \la x\ra \la y\ra \
 \e^{-|x-y|^\zeta} \quad \text{for all} \ x,y \in \Z^2 .
\end{equation}
\emph{(Sufficiently fast polynomial decay would suffice for our purposes.
Note that in the special case
when $E$ is in a spectral gap of $H_{B,\lambda,\omega}$ 
 a simple argument based on the Combes-Thomas estimate  yields
  exponential decay, i.e., $\zeta=1$.)}
\end{quotethm}

\begin{quotethm}[\textbf{SUDEC}]    For $\P$-a.e. $\omega$
the  Hamiltonian
$H_{B,\lambda,\omega}$ has pure point spectrum in
 $\Xi_{B,\lambda}^{\text{DL}}$ with the following property:  Given a closed interval
 $I \subset \Xi_{B,\lambda}^{\text{DL}}$, let
$\{\phi_{n,\omega}\}_{n\in \mathbb{N}}$
 be a complete orthonormal set of  eigenfunctions
of $H_{B,\lambda,\omega}$ with  eigenvalues
$E_{n,\omega}\in I$;  for each $n$ we denote by $P_{n,\omega}$ the 
one-dimensional orthogonal projection on the span of $\phi_{n,\omega}$ and
 set
$\alpha_{n,\omega} =\norm{\la x\ra^{-2}P_{n,\omega}}_2^2= \lVert \la x\ra^{-2}\phi_{n,\omega}  \rVert^2$.  Then
for each  $\zeta \in ]0,1[$
there exists $ C_{I,\zeta,\omega}<\infty$ such that for all $ x,y \in \Z^2$
we have
\begin{equation}\label{decayphi}
\norm{\chi_x P_{n,\omega} \chi_y}_2 =
\lVert \chi_x\phi_{n,\omega}  \rVert\lVert \chi_y\phi_{n,\omega}  \rVert \le
 C_{I,\zeta,\omega} \alpha_{n,\omega} \la x\ra^2 \la y\ra^2
\  \e^{-|x-y|^\zeta} .
\end{equation}
Moreover, we have
\begin{equation}\label{sumalpha}
\sum_{n \in \N} \alpha_{n,\omega} = \mu_\omega(I) :=
\tr \left\{ \la x\ra^{-2} P_{B,\lambda,I,\omega}  \la x\ra^{-2} \right\}< \infty.
\end{equation}
\emph{(Almost-sure pure point spectrum is well known, e.g., 
\cite{FMSS,VDK,GK1,K3}. Property SUDEC, namely \eqref{decayphi} with
\eqref{sumalpha}, is a modification of Germinet's WULE \cite{Ge}.
It is the almost everywhere
consequence (by the Borel-Cantelli Lemma) of a
a new characterization of the region of dynamical 
localization 
given by Germinet and Klein \cite{GK6}.  SUDEC
 is  equivalent to SULE-type properties.)}
\end{quotethm}

\begin{remark2} \emph{Throughout this work we characterize the localization regime
 using consequences of the multiscale analysis.  If the single site bumps of the
Anderson-type potential cover the whole space, i.e. if $\sum_{i\in \Z^d} u(x-i) \ge \delta > 0$, then another
option is available, namely the fractional moment method [AENSS],
which yields exponential bounds for expectations. However at this time the
fractional moment method is not available for potentials which violate the
aforementioned ``covering condition."}
\end{remark2}

We now turn to the Hall conductance.  Consider  the switch function
$\Lambda (t) = \chi_{[\frac 1 2,\infty)}(t)$
and let 
$\Lambda_j$ denotes multiplication by the function $\Lambda_j(x)= \Lambda (x_j)$,
$j=1,2$.  Given an orthogonal projection $P$ on
$\mathrm{L}^2(\mathbb{R}^2, {\mathrm{d}}x)$, we set
 \begin{align}\label{Theta}
\Theta(P):= 
 \tr  \left\{ P\left[  \left[ P,\Lambda_1 \right] , 
\left[ P,\Lambda_2 \right] \right]
\right\},
\end{align}
defined  whenever 
\begin{equation}\label{|Theta|}
|\Theta| (P) :=\norm{  P\left[  \left[ P,\Lambda_1 \right] , 
\left[ P,\Lambda_2 \right] \right]}_1< \infty,
\end{equation}
in which case we also have
 \begin{align}\label{Theta2}
\Theta(P)=  \tr  \left\{ \left[  P\Lambda_1 P, 
 P\Lambda_2P \right] \right\}.
\end{align} 
Note that although $\Theta(P)$ is the trace of a commutator it
need not be zero, because the two summands $P\Lambda_1P \Lambda_2 P$
and $P\Lambda_2 P\Lambda_1P$ are not separately trace class.

\begin{lemma} \label{lemTheta} Let $P$ be an orthogonal projection on
$\mathrm{L}^2(\mathbb{R}^2, {\mathrm{d}}x)$ such that
for some $\xi \in ]0,1]$, $\kappa >0$, and $K_P<\infty$ we have
\begin{equation}\label{decayP}
\norm{\chi_x P \chi_y}_2 \le
K_P \la x\ra^\kappa \la y\ra^\kappa \
 \e^{-|x-y|^\xi} \quad \text{for all} \ x,y \in \Z^2 .
\end{equation}
Then:\\
\noindent{\textbf{(i)}}  $|\Theta| (P) \le C_{\xi,\kappa} K_P^2$ for some constant $C_\xi$
 independent of $P$,
and $\Theta(P)$ is well defined. \\
\noindent{\textbf{(ii)}} Given  $s\in \R$, let
  $\Lambda^{(s)}(t)= 
\Lambda(t-s)$ and $\Lambda_j^{(s)}(x)= \Lambda^{(s)}(x_j)$, $j=1,2$. Set
 $\Theta_{r,s}(P)=  \tr  \left\{ P\left[  \left[ P,\Lambda^{(r)}_1 \right] , 
\left[ P,\Lambda^{(s)}_2 \right] \right]
\right\}$, $r,s \in \R$.
  Then 
 $\Theta_{r,s}(P)$ is well defined as 
in (i), and
$\Theta_{r,s}(P)=\Theta(P)$.\\
\noindent{\textbf{(iii)}} Let $Q$ be another orthogonal projection on
$\mathrm{L}^2(\mathbb{R}^2, {\mathrm{d}}x)$, such that $Q$ commutes with 
$P$ and  also satisfies
\eqref{decayP} with some constant $K_Q$. 
  Then $P+Q$ is an orthogonal projection
satisfying  \eqref{decayP}  with constant $K_{P+Q}=K_P + K_Q$,
and we have 
\begin{equation}
\Theta({P+Q})=\Theta(P) + \Theta(Q).  \label{Thetalinear}
\end{equation}
\end{lemma}

\begin{remark2}   \emph{(i) is similar to statements in \cite{ASS,AG}, 
(ii) and (iii) are well-known \cite{ASS,BES,AG}.  We provide a short proof 
in our
 setting;  the precise form of the bound in \eqref{decayP}
is important for Lemma~\ref{lemconduc}.
Lemma~\ref{lemTheta} remains true if $\Lambda$ is replaced by any
monotone "switch function," with $\Lambda(t) \rightarrow 0,1$ as $t
\rightarrow -\infty, +\infty$, with essentially the  same proof. }
\end{remark2}

\begin{proof}
If $ x \in \Z^2$  we have 
$\Lambda_j \chi_x= \Lambda(x_j ) \chi_x$, $j=1,2$, and hence, if
$x_1  y_1 > 0$ we get
$\chi_x [P,\Lambda_1] \chi_y = 
(\Lambda(y_1)-\Lambda(x_1))\chi_x {P}
\chi_y = 0$.  
If  $x_1y_1 \le 0$,
we have 
$
 |x_1-y_1|^{\xi} \ge  \frac12|x_1|^{\xi} + \frac12|y_1|^{\xi} 
$. Thus it follows from \eqref{decayP}  that
 for all $x,y \in \Z^2$ we have
\begin{align}\label{estcomm}
\norm{\chi_x [{P},\Lambda_1] \chi_y}_2
\le K_P \la x\ra^\kappa  \la y\ra^\kappa \ \mathrm{e}^{- \frac14|x_1|^{\xi} - \frac14|y_1|^{\xi}-
\frac12 |x_2-y_2|^{\xi}  }, 
\end{align}
and, similarly,
\begin{align}
\norm{\chi_x [{P},\Lambda_2] \chi_y}_2
\le  K_P \la x\ra^\kappa  \la y\ra^\kappa  \ \mathrm{e}^{- \frac14|x_2|^{\xi} - \frac14|y_2|^{\xi}-
\frac12 |x_1-y_1|^{\xi}  }.
\end{align}
We conclude  that 
\begin{align} \label{split3factors}
\norm{{P} [{P},\Lambda_1][{P},\Lambda_2]}_1
\le  
\sum_{x,y,z\in\Z^2} \norm{\chi_x [{P},\Lambda_1] \chi_y}_2
\norm{\chi_y [{P},\Lambda_2] \chi_z}_2 \le C_1 K_P^2< \infty , 
\end{align} 
where $C_1$ is a finite constant independent of $P$, and similarly
 $\norm{{P} [{P},\Lambda_2][{P},\Lambda_1]}_1\le C_1  K_P^2$.
Part (i) follows.

 The only nontrivial item in (iii) is \eqref{Thetalinear}.
It follows from \eqref{Theta},  cyclicity of the trace, and
the fact that 
  $P[Q,\Lambda_j] =-P\Lambda_j Q$ for $j=1,2$.

 The proof of (i) clearly applies also
to $\Theta_{r,s}(P)$; we only need to show that $\Theta_{r,s}(P)=\Theta(P)$.
This will follow if we can show that
\begin{equation}\label{F1G2}
\tr  \left\{ P\left[  \left[ P,F_1 \right] , \left[ P,G_2 \right] \right]\right\}
 =\tr  \left\{ P\left[  \left[ P,G_1 \right] , \left[ P,F_2 \right] \right]\right\}
=0,
\end{equation}
if $F = \Lambda^{(s)} -\Lambda^{(s^\prime)}$ and 
$G= \Lambda^{(s^{\prime\prime})}$ for some 
$s,s^\prime, s^{\prime\prime}\in \R$, with
$F_j(x)= F(x_j)$,   $G_j(x)=G(x_j)$, $j=1,2$.  
If we write a trace without a comment, as in \eqref{F1G2},
 we are implicitly stating that
it is well defined by the argument in  \eqref{estcomm}-\eqref{split3factors}.

We have
\begin{align}\notag
&\tr  \left\{ P\left[  \left[ P,F_1 \right] , \left[ P,G_2 \right] \right]\right\}=
\tr  \left\{ P F_1  (I - P) \left[ P,G_2 \right]\right\}+ 
\tr  \left\{\left[ P,G_2 \right] (I - P) F_1P\right\}\\ 
&\qquad = \tr \left\{ F_1  (I - P) \left[ P,G_2 \right] P +
F_1P  \left[ P,G_2 \right] (I - P)    \right\}\\
&\qquad = \tr \left\{ F_1  \left[ P,G_2 \right]    \right\}
=  \tr \left\{  \left[ F_1 P,G_2 \right]    \right\}.\notag
\end{align}
Here $F_1\left[ P,G_2 \right] =\left[ F_1 P,G_2 \right]  $ is trace class as the sum of
 two trace class operators.  This can also be seen directly as follows:  The
function $F_1(x)$ has compact support in the $x_1$ direction, and using the fact that $P$ is a 
projection
satisfying \eqref{decayP} we get
\begin{equation}\label{decayPinT1}
\norm{\chi_x P \chi_y}_1 \le
C{ K^2_P} \la x\ra^{2\kappa} \la y\ra^{2\kappa} \
 \e^{-\frac 12  |x-y|^\xi} \quad \text{for all} \ x,y \in \Z^2 ,
\end{equation}
for some constant $C $.  Since $F_1PG_2$ and $G_2 F_1P$ may 
not be trace class, we introduce a cutoff
 $Y_n (x)=\chi_{[-n,n]}(x_2)$ in the $x_2$ direction.  Note
\begin{equation}
 \tr \left\{  Y_n \left[ F_1 P,G_2 \right]\right\} =
 \tr \left\{  \left[ Y_n F_1 P,G_2 \right]\right\}=0,
\end{equation}
since  $Y_n F_1PG_2$ and $Y_n G_2 F_1P$ are trace class
by \eqref{decayPinT1} and the argument in the  
proof of Lemma~\ref{lemTheta}.  Thus
 \begin{equation}
\tr \left\{  \left[F_1  P,G_2 \right]\right\}= 
\lim_{n\to\infty} \tr \left\{  Y_n \left[ F_1 P,G_2 \right]\right\}=0.
\end{equation}

The other term in \eqref{F1G2} is treated in the same way, and Part (ii) is proven.
\end{proof}

For a given disorder  $\lambda\ge 0$,   magnetic field $B>0$, and energy 
 $E \in\Xi_{B,\lambda}^{\text{DL}}$,
 we consider the Hall conductance \cite{ASS,ES} 
\begin{align}\label{sigmaomega}
\sigma_{H,\omega} (B,\lambda,E)= - 2\pi i\,
 \Theta (P_{B,\lambda,E,\omega}) ,
\end{align}
well defined  for $\P$-a.e.\ $\omega$
in view of Lemma~\ref{lemTheta}(i) and  (DFP), namely \eqref{decayFBC}.
The covariance relation  \eqref{covariance}  and Lemma~\ref{lemTheta}(ii)
 then imply
 $\sigma_{H,\omega}(B,\lambda,E)= \sigma_{H,\tau_a\omega}(B,\lambda,E)$
for all $a \in \Z^2$ for $\P$-a.e.\  $\omega$, and hence 
 ergodicity yields
\begin{equation}\label{sigmaerg}
\sigma_H(B,\lambda,E) := \E \left\{\sigma_{H,\omega} (B,\lambda,E) \right\}
=\sigma_{H,\omega}(B,\lambda,E) \quad 
\text{for $\P$-a.e.\ $\omega$}. 
\end{equation}

A key ingredient in justifications of the quantum Hall effect  is that
the Hall conductance should be constant in intervals of localization
since localized states do not carry current \cite{L,Ha,Ku,BES}.  The following lemma
 makes this precise in a very transparent way: In intervals of dynamical
localization the change in the Hall conductance is given by the (infinite) sum 
of the Hall conductance of the individual localized states.  But the conductance
of a localized state is equal to $-2 \pi i\Theta(P)$, where $P$ is the orthogonal
projection on the localized state.  But if  $P$ is a  one-dimensional orthogonal
projection, say on the span of unit vector $\psi$,  \eqref{Theta2} yields
\begin{equation}\label{Theta=0}
\Theta(P)= \la \psi, \Lambda_1 \psi\ra  \la \psi, \Lambda_2 \psi\ra -
\la \psi, \Lambda_2 \psi\ra \la \psi, \Lambda_1 \psi\ra=0.
\end{equation}

\begin{lemma}\label{lemconduc} The Hall conductance
$\sigma_H(B,\lambda,E)$ is constant on connected components 
of $\Xi_{B,\lambda}^{\text{DL}}$, that is, if 
 $[E_1,E_2]\subset \Xi_{B,\lambda}^{\text{DL}}$ we must have
 $\sigma_H(B,\lambda,E_1)=\sigma_H(B,\lambda,E_2)$.
\end{lemma}

\begin{proof} 

If  $I=[E_1,E_2]\subset \Xi_{B,\lambda}^{\text{DL}}$,   we apply 
property
 (SUDEC) in $I$
for the $\P$-a.e.\ $\omega$ for which we have
\eqref{decayphi} and \eqref{sumalpha}. Given a (finite or infinite)
 subset $M$ of the index set 
$\N$, we set $P_{M,\omega} = \sum_{n \in M} P_{n,\omega}$; it follows 
that we have condition \eqref{decayP} for $P_{M,\omega}$ for $\kappa=2$
and  all
$\zeta \in ]0,1[$ with constant 
\begin{equation}\label{KPM}
K_{P_{M,\omega}}=C_{I,\zeta,\omega} \sum_{n \in M}  \alpha_{n,\omega}
\le C_{I,\zeta,\omega} \,\mu_\omega(I)< \infty.
\end{equation}

Since
$ P_{B,\lambda,]E_1,E_2],\omega} =P_{B,\lambda,E_2,\omega}
-P_{B,\lambda,E_1,\omega}$, it follows from 
Lemma~\ref{lemTheta}, (i) and (iii),
 that it suffices to prove that
\begin{equation}
\Theta( P_{B,\lambda,]E_1,E_2],\omega})=0.
\end{equation}
But again using Lemma~\ref{lemTheta}, (i) and (iii),
taking $M=\{1,2 \dots,m\} \subset \N$, we have 
\begin{align}
\Theta( P_{B,\lambda,]E_1,E_2],\omega})&= 
\Theta(P_{\N,\omega})= \Theta(P_{ M,\omega})+
  \Theta(P_{(\N\backslash M),\omega})\\
&= \sum_{n=1}^m  \Theta(P_{ n,\omega}) 
+\Theta(P_{(\N\backslash M),\omega}) \notag.
\end{align}
Since by Lemma~\ref{lemTheta}(i), \eqref{KPM} and \eqref{sumalpha} we have
\begin{equation}
\left\vert\Theta(P_{(\N\backslash M),\omega})\right\rvert \le 
C_\zeta\left(C_{I,\zeta,\omega}
 \sum_{n=m+1}^\infty \alpha_{n,\omega}\right)^2 \to 0 \quad 
\text{as $m \to \infty$},
\end{equation}
we conclude that
\begin{equation}
\Theta( P_{B,\lambda,]E_1,E_2],\omega}) = 
\sum_{n=1}^\infty \Theta(P_{ n,\omega})=0
\end{equation}
in view of \eqref{Theta=0}.
\end{proof}

\begin{remark2}\emph{ The constancy of the Hall conductance
in intervals of localization is known for lattice
Hamiltonians with eigenvalues of finite multiplicity [BES, AG,EGS].
Since do not rule out eigenvalues of 
infinite multiplicity  for random Landau Hamiltonians,
 we must take subsets of
eigenfunctions, not of the interval $I$.  The crucial estimate is thus 
\eqref{KPM}, a consequence of property (SUDEC). }
\end{remark2}

In the  next lemma, we  calculate the value of the Hall conductance in the spectral 
gaps between the  bands
under the disjoint bands condition.

\begin{lemma}\label{lemconducgap} 
 Under the disjoint bands conditions \eqref{gapcond} we have 
$\sigma_H(B,\lambda,E)=n$ if $E\in  \mathcal{G}_n(B,\lambda)$ for
 all $n=0,1,2\dots$.
\end{lemma}

\begin{proof}
 It is well known that 
 $\sigma_H(B,0,E)=n$ if $E\in]B_{n},B_{n+1}[$ for all $n=0,1,2\dots$ 
\cite{ASS,BES}.
Under condition \eqref{gapcond}, if $E\in  \mathcal{G}_n(B,\lambda_1)$ for
 some $n\in \{0,1,2\dots\}$ we can find $\lambda_E > \lambda_1$ such that
$E\in  \mathcal{G}_n(B,\lambda)$ for all $\lambda \in I= [0,\lambda_E[$.
 We take $\omega \in   [-M_1,M_2]^{\Z^2}$,
a set of probability one, and
  note that the contour $\Gamma$ below and all the constants on what follows
 are independent of $\omega$.  We have
\begin{equation}\label{contourint}
P_\lambda = - \tfrac 1 {2\pi i} \int_\Gamma R_\lambda(z) \,\di z  \quad 
\text{for all} \ \lambda \in I,
\end{equation}
where $P_\lambda = P_{B,\lambda,E,\omega}  $,
 $R_\lambda(z) = (H_{B,\lambda,\omega}-z)^{-1}$, and $\Gamma$ is a
 bounded contour such that  $d(\Gamma,\sigma (H_{B,\lambda,\omega})) \ge \eta >0$
for all $\lambda \in I$.
 (Note $H_{B,\lambda,\omega} \ge B- \lambda_E M_1$ for all 
$\lambda \in I$.)  We have ($K_1,K_2,\dots$ denote constants) 
\begin{align}\label{CT}
\|\chi_x R_\lambda(z) \chi_y\| &\le K_1 \e^{-K_1 |x-y|} \quad \text{for all} \ x,y \in \Z^2,
z \in \Gamma, \lambda \in I ,\\
\label{HSest}
\| R_\lambda(z) \chi_x\|_2 &\le K_2 \ \quad \text{for all} \ x \in \Z^2,
z \in \Gamma, \lambda \in I ,
\end{align}
where \eqref{CT} is the Combes-Thomas estimate
(e.g., \cite[Corollary 1]{GK2}) and  \eqref{HSest} is in
\cite[Proposition 2.1]{BGKS}.  Combining with \eqref{contourint}, we get
\begin{align}\label{CT1}
\|\chi_x P_\lambda \chi_y\| &\le \tfrac {K_1 |\Gamma|}{2 \pi} \, \e^{-K_1|x-y|} 
\quad \text{for all} \ x,y \in \Z^2,\lambda \in I ,\\
\label{trkernel}
\|\chi_x P_\lambda \chi_y\|_1 &\le \left(\tfrac {K_2 |\Gamma|}{2 \pi}\right)^2
 \quad \text{for all} \ x,y \in \Z^2,\lambda \in I ,\\
\label{CTHS}
\|\chi_x P_\lambda \chi_y\|_2 &\le K_3\, \e^{-K_3 |x-y|} 
\quad \text{for all} \ x,y \in \Z^2,\lambda \in I ,
\end{align}
where \eqref{CTHS} follows from  \eqref{CT1} and \eqref{trkernel}.

Given  $\lambda,\xi\in I$, it follows from \eqref{contourint} and the resolvent 
identity that
\begin{align}\label{taylor}
Q_{\lambda,\xi}:= P_{\xi} -P_\lambda =
\tfrac {(\xi -\lambda)} {2\pi i} \int_\Gamma R_\lambda(z) V R_\xi (z)\,\di z,
\end{align}
with $V= V_\omega$ (recall $\|V\| \le \max \{M_1,M_2\}$). 
Using \eqref{CT} and \eqref{HSest}, we get
\begin{equation}\label{CTHS2}
\|\chi_x Q_{\lambda,\xi} \chi_y\|_2 \le K_4\, \e^{-K_4 |x-y|}
\quad \text{for all} \ x,y \in \Z^2,\lambda,\xi \in I. 
 \end{equation}

We now use an idea of
Elgart and Schlein \cite{ES}.
If the potential $V$ had compact support, it would follow from
\eqref{HSest} that  $Q_{\lambda,\xi}$ is trace class.  In this case,
using \eqref{Theta2} and \eqref{taylor}, we get
\begin{align}\label{eqcomp}
\Theta( P_{\xi}) -  \Theta( P_{\lambda}) &=
\tr  \left\{ \left[  Q_{\lambda,\xi}\Lambda_1  P_{\xi}, 
  P_{\xi}\Lambda_2 P_{\xi}\right]  + 
\left[   P_{\lambda}\Lambda_1 Q_{\lambda,\xi}, 
  P_{\xi}\Lambda_2 P_{\xi} \right]  \right.\\
& \qquad \left.+
 \left[  P_{\lambda}\Lambda_1 P_{\lambda}, 
 Q_{\lambda,\xi}\Lambda_2 P_{\xi} \right] + 
\left[  P_{\lambda}\Lambda_1 P_{\lambda}, 
 P_{\lambda}\Lambda_2Q_{\lambda,\xi} \right]\right\}=0 ,\notag
\end{align}
since $\tr [A,B]=0$ if either $A$ or $B$ are trace class by 
centrality of the trace.  Our potential $V$, given in \eqref{potVL},
does not have compact support, so we will use an approximation argument.

Given $L >0$ and
$\omega \in   [-M_1,M_2]^{\Z^2}$, we define
 $\omega^{(L)}, \omega^{(>L)}\in   [-M_1,M_2]^{\Z^2}$ by
$\omega^{(L)}_i = \omega_i$ if $ |i| \le L$ and $\omega^{(L)}_i =0$ otherwise.
and $ \omega^{(>L)}_i =\omega_i - \omega^{(L)}_i$ for all $i \in \Z^2$.
Recalling \eqref{potVL}, we set $V_L= V_{\omega^{(L)}}$, 
$V_{>L}= V_{\omega^{(>L)}}= V-V_L$,
 $P_{\lambda,L} = P_{B,\lambda,E,\omega^{(L)}}  $, 
$P_{\lambda,>L} = P_{B,\lambda,E,\omega^{(>L)}}= P_{\lambda}-P_{\lambda,L} $, 
etc.  We have
\begin{align}\label{taylor2}
Q_{\lambda,>L}:= P_{\lambda} - P_{\lambda,L} = 
\tfrac {\lambda} {2\pi i} 
\int_\Gamma R_\lambda(z) V_{>L}R_{\lambda,L}  (z)\,\di z. 
\end{align}
Moreover, it follows from \eqref{taylor2} and \eqref{CT} that
\begin{equation}\label{CT1'}
\|\chi_x Q_{\lambda,>L} \chi_y\| \le K_5
\, \e^{-K_5 (\max\{L-|x|,0\}+\max\{L-|y|,0\}} \, \e^{-K_5 |x-y|}
\end{equation}
for all $x,y \in \Z^2,\lambda \in I$ and $L>0$, with a similar estimate holding
in the Hilbert-Schmidt norm by the argument used for \eqref{CTHS}. 
Using \eqref{Theta} and \eqref{taylor2}, we have
\begin{align}\notag
&\Theta( P_{\lambda}) -  \Theta( P_{\lambda,L}) = 
\tr  \left\{ Q_{\lambda,>L}\left[  \left[ P_{\lambda},\Lambda_1 \right] , 
\left[ P_{\lambda},\Lambda_2 \right]\right] + P_{\lambda,L}\left[  \left[ Q_{\lambda,>L},\Lambda_1 \right] , 
\left[ P_{\lambda},\Lambda_2 \right]\right] \right. \\
&\left. \qquad  \qquad+
 P_{\lambda,L}\left[  \left[ P_{\lambda,L},\Lambda_1 \right] , 
\left[ Q_{\lambda,>L},\Lambda_2 \right]\right]
\right\} \to 0 \quad \text{as $L \to \infty$},\label{to0}
\end{align}
where the convergence to $0$ is proved as follows: 
Since $\| Q_{\lambda,>L}\| \le  K_6$ for all $L >0$ and 
$ Q_{\lambda,>L} \to 0$ strongly as $L \to \infty$, the trace of the first term
goes to $0$ as $L \to \infty$.  The traces of the other two terms 
can be estimated as in \eqref{split3factors}, and 
converge to $0$ as $L \to \infty$ by an argument using 
 \eqref{CT1'} and dominated convergence.

The lemma now follows from \eqref{eqcomp} and \eqref{to0}.
\end{proof}

We may now finish the proof of Theorem~\ref{corbeta}. 
Since \eqref{gapcond} holds, if
 $\mathcal{B}_n(B,\lambda)\subset \Xi_{B,\lambda}^{\text{DL}} $
for some $n\in \{1,2,\dotsc\}$ we have
 $$] B_{n-1} +\lambda M_1,B_{n+1} - \lambda M_2 [  
=\mathcal{G}_{n-1}(B,\lambda) \cup \mathcal{B}_n(B,\lambda) 
\cup   \mathcal{G}_n(B,\lambda)\subset \Xi_{B,\lambda}^{\text{DL}},$$
and hence  it follows from Lemma~\ref{lemconduc} that the Hall
 conductance 
$\sigma_{H} (B,\lambda,E)$ has the same value on the spectral gaps
$\mathcal{G}_{n-1}(B,\lambda)$ and $\mathcal{G}_{n}(B,\lambda)$,
which contradicts  Lemma~\ref{lemconducgap}.  Thus we must have
 $\mathcal{B}_n(B,\lambda)\cap \Xi_{B,\lambda}^{\text{DD}}\not=\emptyset $
for all $n\in \{1,2,\dotsc\}$, and hence   Theorem~\ref{corbeta}
follows from property (RDD).

\section{The applicability of the multiscale analysis}
\label{sectWegner}

In order to use properties (RDL), (RDD), (DFP), and (SUDEC), we must show that
the random Landau Hamiltonian  $H_{B,\lambda,\omega}$ ($\lambda > 0$)
  satisfy the hypotheses in  \cite{GK1,GK5} at all energies,
including the Landau levels. These were called  assumptions or properties
SGEE, SLI, EDI, IAD, NE, and W  in \cite{GK1,GK3,GK5,K3}.  (Although
the results in  \cite{GK1,GK5} are written for random Schr\"odinger operators
without magnetic fields, they hold without change with 
 magnetic fields as long as these hypotheses are satisfied.)

Property SGEE guarantees the existence
of   a  generalized eigenfunction expansion in the strong sense (the required
trace  estimate holds in expectation) and is known for a large class
of random operators which includes the random Landau Hamiltonian
(the trace estimate for  Schr\"odinger operators with magnetic
fields can be found in \cite[Proposition 2.1]{BGKS}).

Properties SLI, EDI, IAD, NE, and W are {the requirements for 
 a multiscale analysis}, and are properties concerning an appropriate
 finite volume  restriction of the random Schr\"odinger operator.
 For the random Landau Hamiltonian the finite volumes may be
the squares $ \Lambda_L(x)$
with $x \in \Z^2$ and $L\in L_0\N$ for a suitable $L_0 \ge 1$.
The multiscale analysis requires the notion of a 
{\em finite volume operator}, a ``restriction"
$H_{B,\lambda,\omega,x,L}$ of $H_{B,\lambda,\omega}$ to the square
$\Lambda_L(x)$  where the ``randomness based
 outside the square  $\Lambda_L(x)$" is not taken into account. 
Usually the finite volume operator is defined as an 
operator on $\mathrm{L}^2(\Lambda_L(x),{\rm d}x)$ by specifying
the boundary condition, most commonly Dirichlet or periodic boundary
condition. (In the case of the random Landau Hamiltonian it has also been
defined as an operator  on the whole space by throwing away the random coefficients ``based
 outside the square $\Lambda_L(x)$" \cite{CH2,Wa,GK4}.)

\emph{But it is not necessary to use the same boundary condition on 
all squares.}
For the multiscale analysis it suffices to fix a scale  $L_0 \ge 1$, \emph{not
necessarily an integer}, fix some  $\varrho>0$, and define
a random operator  $H_{B,\lambda,\omega,x,L} $  on
 $\mathrm{L}^2(\Lambda_L(x),{\rm d}x)$ for each
  $x \in \Z^2$ and $L\in L_0\N$  as follows: First pick
  a closed densely defined
operator $ \Db_{B,x,L}$ from $\mathrm{L}^2(\Lambda_L(x),{\rm d}x)$ to
$\mathrm{L}^2(\Lambda_L(x),{\rm d}x; \C^2)$ which is an extension
of the differential operator   $\Db_B= (-i\nabla-\mathbf{A})  $
restricted to $C^\infty_c (\Lambda_L(x))$.  Second,  pick
  a  random potential $ V_{x,L,\omega}$
 in the square $\Lambda_L(x)$ depending only
on the random variables $\{\omega_i; \, i \in \Lambda_L(x)\}  $,
and set
$H_{B,\lambda,\omega,x,L}=
 \Db_{B,x,L}^* \Db_{B,x,L} + \lambda  V_{x,L,\omega} $  on
 $\mathrm{L}^2(\Lambda_L(x),{\rm d}x)$. 
 Require of the operators $\Db_{B,x,L}$ that the resulting 
 $H_{B,\lambda,\omega,x,L}$ have  compact resolvent  and satisfy the
 covariance condition (\emph{but only between $x$ and $0$, not between arbitrary
 $x$ and $y$ in $\Z^2$}) 
\begin{equation}\label{covL}
 H_{B,\lambda,\omega, x,L}= U_x H_{B,\lambda,\tau_{-x}(\omega),0,L} U_x^* 
\quad \mbox{for  all $ x \in \mathbb{Z}^2$},
\end{equation}
where the magnetic translation $U_x$ is as in  \eqref{magtrans}
but considered as a unitary map from 
$\mathrm{L}^2(\Lambda_L(0),{\rm d}x)$ to
 $\mathrm{L}^2(\Lambda_L(x),{\rm d}x)$.   Furthermore,
require the following  compatibility
conditions:  If
$\vphi \in \D( \Db_{B,x,L})  $ with
 $\supp \vphi \subset  \Lambda_{L-\varrho}(0)$, then
$\I_L\vphi \in \D(\Db_{B})  $, and  
\begin{equation}\label{compatibility}
\mathcal{I}_{L} \Db_{B,x,L} \varphi =
\Db_{B} \mathcal{I}_{L}\vphi,\quad 
 \mathcal{I}_{L}\chi_{x, L-\varrho}  V_{x,L,\omega}  =
\chi_{x, L-\varrho} V_{\omega},
 \end{equation}
 where $\mathcal{I}_{L}\colon \mathrm{L}^2(\Lambda_L(0),{\rm d}x)
\to \mathrm{L}^2(\R^2,{\rm d}x)$  is the canonical injection:
$\lt(\mathcal{I}_{L}\vphi\rt) (x)=\vphi(x) $ if  $ x \in\Lambda_L(0)$,
$\lt(\mathcal{I}_{L}\vphi\rt) (x)=0 $ otherwise (we  also use 
$\mathcal{I}_{L}$
for $\C^2$ valued functions). 
This  is equivalent to  fixing  the boundary condition for
 the operators $ \Db_{B,x,L}$
  at the square centered at
$0$, and using the magnetic translations to define the finite volume
operators in all other squares by \eqref{covL}; note that in the square
centered at  $ x \in \Z^2$ with side $L -\varrho$ the potential 
 $V_{x,L,\omega}$ is just $V_\omega$.  (This also
applies for ``finite volume operators"  defined on the whole space,
except that these operators are only relatively compact perturbations 
of  $H_B$.)

 One must then 
show that the properties
SLI, EDI, IAD, NE, and W  hold for these finite volume operators.
Only properties W (the Wegner estimate) and 
NE (average number of eigenvalues) present difficulties. 
Property IAD (independence at a distance) is  obvious. 
Properties SLI (Simon-Lieb inequality) and EDI
 (eigenfunction decay inequality) follow from \eqref{covL} 
and \eqref{compatibility} as in  \cite[Theorem A.1]{GK5}
(see  also the discussion
in \cite[Section 4]{GK3}).

If the single bump potential $u$ in $\eqref{potVL}$ has $\eps_u \ge 1$,
then properties W and NE
are proven for appropriate finite dimensional operators in \cite{HLMW} at all energies.  
 But if  $\eps_u$ is small (the most
 interesting case for this paper in view of Corollary~\ref{corlimit1}), a Wegner
 estimate (and Assumption NE)
at all energies was only known under the rational flux condition on the unit
square, namely
$B \in 2\pi \mathbb{Q}$  \cite{CHK};
otherwise
a Wegner estimate was  known under  the hypotheses of
Corollary~\ref{corlimit1}
but only at  energies different
 from the Landau levels \cite{CH2,Wa}.

The Wegner estimate is closely connected to H\"older continuity of the
integrated density of states, in fact Combes, Hislop and Klopp 
\cite{CHK} proved first a Wegner estimate for random Landau Hamiltonians
with $B \in 2\pi \mathbb{Q}$, and from it derived the
 H\"older continuity of the integrated density of states.
 Combes, Hislop, Klopp and Raikov \cite{CHKR} 
established the   H\"older continuity of the
integrated density of states for $H_{B,\lambda,\omega}$ as in \eqref{landauh}
 with no extra hypotheses, but they
 did not obtain estimates on  finite volume operators, and
 hence no Wegner estimate.

In the next theorem we establish a Wegner estimate (and also property NE)
 for the random
Landau Hamiltonian as in \eqref{landauh}, for an appropriate choice of
 finite volume  operators.  Although the Wegner estimate does not
follow from H\"older continuity of the
integrated density of states,  we  use some of the key results in 
 \cite{CHKR} to obtain the crucial  estimate \cite[Eq. (3.1)]{CHK}, 
from which the Wegner
estimate follows as in \cite[Proof of Theorem 1.2]{CHK}. 

Let $B >0$ be arbitrary; since we do not assume the rational flux condition on the
 unit square, we  set a length scale corresponding to squares
 with even (for convenience) integer flux.
We take $K_B= \min \Bigl\{k \in \N; k \ge \sqrt{\tfrac B {4\pi}}\Bigr\}$,
 and set
\begin{equation}\label{LB}
L_B= K_B \sqrt{\tfrac{4\pi}  B } , \quad 
\N_B  = L_B \N, \quad \text{and} \quad   \Z^2_B  = L_B \Z^2.
\end{equation}
Note that $L_B\ge 1$ may not be an integer. We consider squares
$\Lambda_L(0)$ with $L \in \N_B$ and identify them with the torii
$\mathbb{T}_L:=\R^2/(L\Z^2)$
in the usual way.  As shown
 in \cite[Section 4]{CHK}, the magnetic translations 
$\mathcal{U}_{{B}}:=\{U_a; \ a \in {\Z^2_B}\}$ form a unitary representation 
of the abelian group $\Z^2_B$; we write $\widehat{U}_a$ for the 
corresponding action on  $\mathrm{L}^2(\Lambda_L(0),{\rm d}x)$, with  
$\widehat{\mathcal{U}}_{{B}}:=\{\widehat{U}_a; \ a \in {\Z^2_B}\}$. If 
$x \in \Lambda_L(0)$ and $r <L$
we denote by $\widehat{\Lambda}_r(x)$ and $\widehat{\chi}_{x,L}$
the square and characteristic function in the
 torus $\mathbb{T}_L$.

Given $L \in{{\N_B}}$,
we define $H_{B,0,L}=\Db_{B,0,L}^* \Db_{B,0,L}$, with $\Db_{B,0,L}$
 the restriction of $\Db_B$ to
$ \mathrm{L}^2(\Lambda_L(0), {\mathrm{d}}x)$ with periodic boundary 
condition
with respect to $\widehat{\mathcal{U}}_{{B}}$.  The spectrum of $H_{B,0,L}$
still consists of the Landau levels:
$\sigma(H_{B,0,L})= \sigma(H_{B})=\{ B_n; \, n=0,1,\ldots\}$,
but since $H_{B,0,L}$ has compact resolvent each Landau level
 has now finite multiplicity.  
We let $\widetilde{\Lambda}_L(x)= \Z^2 \cap \Lambda_L(x)$. Given
$L \in {{\N_B}}$ we set
\begin{equation}\begin{split}
 \label{landauh2} 
H_{B,\lambda,0,L,\omega}& = H_{B,0,L} +
\lambda  V_{0,L,\omega} \quad \mathrm{on} \quad
 \mathrm{L}^2(\Lambda_L(0), {\mathrm{d}}x),\\
V_{0,L,\omega}(x)&=  \sum_{i\in  \widetilde{\Lambda}_{L-\delta_u}(0)}
\omega_i \,u(x-i),
\end{split}\end{equation}
where $\supp u \subset \Lambda_{\delta_u}(0)$, and then define  $H_{B,\lambda,\omega,x,L}$ for all $ x \in \Z^2$ 
 by  \eqref{covL}.  (We prescribed periodic boundary condition for the
(free) Landau Hamiltonian  at the square centered at
$0$, and used the magnetic translations to define the finite volume
operators in all other squares by \eqref{covL}; in the square
centered at  $ x \in \Z^2$ the potential  $V_{x,L,\omega}$ is exactly as 
in \eqref{landauh2} except that the sum is now over 
$i\in  \widetilde{\Lambda}_{L-\delta_u}(x) $.) 
Note that $H_{B,\lambda,x,L,\omega}$ has compact resolvent and satisfies
the compatibility conditions
\eqref{compatibility}  for an appropriate $\varrho>0$.

The following theorem establishes both property W and NE for these
finite volume operators at all energies.  We write
$P_{B,\lambda,\omega,x,L}(\mathcal{J}) =
\chi_{\mathcal{J}}(H_{B,\lambda,\omega,x,L})$ if 
 $\mathcal{J}\subset \R$ is a Borel set. Recall that $\rho$ is the bounded density
of the common  probability distribution of the $\omega_i$'s.

\begin{theorem}\label{thmWegner}  Fix $B>0$ and $\lambda >0$.
Given  a bounded interval $I\subset\R$ and $q\in]0,1[$, there exist 
constants $Q_{B,\lambda,I,q}<\infty$ and $\eta_{B,\lambda,I}\in]0,1]$, and a
 finite scale $L_{B,\lambda,I,q}$, such that for all subintervals $J\subset I$
 with $|J|\le \eta_{B,\lambda,I}$,  $L\in {{\N_B}}$ with
$L\ge   L_{B,\lambda,I,q}$, and $ x\in \Z^2$, we have
\begin{equation}\label{ineqWegner}
\E \left\{ \tr P_{B,\lambda,\omega,x,L}(J) \right\} \le
Q_{B,\lambda,I,q} \norm{\rho}_\infty |J|^q L^2.
\end{equation}
\end{theorem}

\begin{proof} In view of \eqref{covL} it suffices to prove the theorem for $x=0$.

 We start by proving a
 lemma that will allow us to derive the theorem from the results
of \cite{CHKR,CHK}.  For each $L \in\N_B$ we set 
$\Gamma_L =
\chi_{ \overline{\Lambda}_{L-1}(0)\backslash \Lambda_{L-3}(0)}$
and  fix a function 
$\Phi_L \in C^\infty (\R^2)$ such that
$\Phi_L(x) \equiv 1$ on $\Lambda_{L-\frac 5 2}(0)$, 
$\supp \Phi_L \subset \Lambda_{L-\frac 3 2}(0)$, and  $0\le\Phi_L(x) \le1$,
$\abs{\nabla \Phi_L(x) } \le 5$ for all $x \in \R^2$. 
(Such a function always exists.)  We   use $\Phi_L$, $(\nabla \Phi_L)$, and $\chi_r = \chi_{0,r}$
($0<r \le L$) to denote the
operators given by multiplication by the functions $\Phi_L$, 
$\nabla \Phi_L $ and
$\chi_r$ in both
 $ \mathrm{L}^2(\Lambda_L(0),{\rm d}x)$ and
$\mathrm{L}^2(\R^2,{\rm d}x)$. For convenience we set
$H_{B,L}=H_{B,0,L}$,
$\widetilde{\N}_B={\N_B} \cup\{\infty\}$,  $H_{B,\infty}=H_{B}$, and so on.
 By  $C_{a,b, \ldots}$ 
we denote a  constant depending only on the
  parameters $a,b,\ldots$ (we may use the same $C_{a,b, \ldots}$ 
for different constants), and similarly for constants 
 $m_{a,b, \ldots}>0$.

\begin{lemma} \label{lemcrucial}Fix $B>0$. Given $n\in \N $ and 
$L\in {\widetilde{\N}_B}$, let
 $\Pi_{n,L}=\Pi_{B,n,L}$  denote the  orthogonal
 projection on the eigenspace corresponding to the $n$-th Landau level
$B_n$ for the  Landau Hamiltonian  $H_{B,L}$. 
 Then for all  $x \in \Lambda_{L_B}(0)$, $r>0$, and  $L\in {{\N}_B}$  
such that $L \ge 2(L_B + r)$,
   we have
\begin{align}
\Pi_{n,L}{\chi_{x,r}}\Pi_{n,L} = \Phi_L\I_L^*\Pi_{n}{\chi_{x,r}} \Pi_{n} \I_L \Phi_L + 
\mathcal{E}_{x,r,n,L},\label{PiLdouble}
\end{align}
 with the error operator $ \mathcal{E}_{x,r,n,L}$ satisfying
\begin{equation}
\norm{\mathcal{E}_{x,r,n,L}} \le   C_{n,B} \,\e^{-m_{n,B} L} . \label{PiL2}
\end{equation}
\end{lemma}

\begin{proof}  Let $L$, $r$, and $x$ be as in the lemma.
Since all $ H_{B,L}$  have the same spectrum, namely the Landau levels,
we have 
\begin{equation}\label{contour}
\Pi_{n,L} =  - \tfrac 1 {2\pi i} \int_{\gamma_n} R_L (z) \,\di z  \
\text{with   $R_L(z) = (H_{B,L}-z)^{-1}$ if $L \in \widetilde{\N}_B$},
\end{equation}
where  $\gamma_n$ denotes the circle centered at $B_n$ with radius $B$.
Let  $z \in \gamma_n$, in view of \eqref{compatibility} we may use the smooth resolvent identity as 
in \cite[Eq. (6.13)]{GK5}
to obtain, 
\begin{equation}\label{SRE}
\begin{split}
{\chi_{x,r}}\I_L  R_L(z)& ={\chi_{x,r}}\Phi_L \I_L  R_L(z)= {\chi_{x,r}}R(z)\Phi_L\I_L
  - {\chi_{x,r}}Y_L(z),
\\
  Y_L(z)& := i   R(z) \lt \{ \Db_B^*(\nabla\Phi) \I_L +
\I_L (\nabla \Phi)^* \Db_{B,L} \rt \} R_L(z) .
\end{split}
\end{equation}
Proceeding as in \cite[Proof of Lemma 6.4]{GK5}, 
using $L \ge 2(L_B + r)$, 
 $\norm{R_L(z)}= \frac 1 B$,  $\abs{z} \le B_n + B$, 
and the Combes-Thomas estimate (e.g., \cite[Theorem 1]{GK2}), we obtain
\begin{align}\notag
\norm{ {\chi_{x,r}}Y_L(z)}&\le
 \norm{ {\chi_{x,r}}R(z)\Db_B^* \abs{\nabla\Phi}}\norm{R_L(z)} 
+\norm{ {\chi_{x,r}}R(z)\abs{\nabla\Phi}} \norm{ \Db_{B,L}  R_L(z)}\\
&\le C_{n,B} \norm{ {\chi_{x,r}}R(z)\Gamma_L}
\le C_{n,B} \, \e^{-m_{n,B} L}.\label{YLz}
\end{align}
Putting together \eqref{contour}, \eqref{SRE}, and  \eqref{YLz} we get
\begin{equation}\label{PiL}
{\chi_{x,r}}\Pi_{n,L} = {\chi_{x,r}}\I_L^* \Pi_{n} \I_L \Phi_L + 
\mathcal{E}^\prime_{x,r,n,L},
\end{equation}
with the error operator $\mathcal{E}^\prime_{x,r,n,L}$ satisfying
the estimate \eqref{PiL2}.  The lemma now follows from 
\eqref{PiL}.
\end{proof}

Using Lemma~\ref{lemcrucial}  we  adapt the crucial
\cite[Lemma~2]{CHKR} to finite volume.

\begin{lemma}\label{lemWegner}Fix $B>0$,  $n\in\N$, $\eps>0$,
 $R>\eps$, and $\eta>0$.  If   $\kappa>1$
and  $L\in \N_B$ are such that $ L >2(L_B  + \kappa R)$, then for all
 $x\in\Lambda_{L}(0)$  we have
\begin{equation}\label{boundlem}
\Pi_{n,L} \widehat{\chi}_{{x,\eps}} \Pi_{n,L} \ge C_0
 \Pi_{n,L} (\widehat{\chi}_{{x,R}} -
\eta\widehat{\chi}_{{x,\kappa R}})\Pi_{n,L} + \Pi_{n,L} \Ec_{n,x,L}
\Pi_{n,L},
\end{equation}
where  $C_0=C_{0;n,B,\eps,R,\eta }>0$ is a constant and       
the error operator $\Ec_{n,x,L}=\Ec_{n,x,L,B,\eps,R,\eta} $
satisfies 
\begin{equation}
\norm{\Ec_{n,x,L}} \le   C_{n,B,\eps,R,\eta} \,\e^{-m_{n,B} L} . \label{PiL3}
\end{equation}
\end{lemma}

\begin{proof}  Given $B,n,\eps, R,\eta$ as in the Lemma, it follows from
\cite[Lemma~2]{CHKR} that for all $\kappa >1$ and $x \in \R^2$ we have
\begin{equation}\label{CHKRbound}
\Pi_{n} \chi_{{x,\eps}} \Pi_{n} \ge C_0
 \Pi_{n} (\chi_{{x,R}} -
\eta\chi_{{x,\kappa R}})\Pi_{n}, \quad  
C_0=C_{0;B,n,\eps,R,\eta}>0.
\end{equation}
(Although \cite[Eq 61]{CHKR} is stated for discs instead of squares, 
\eqref{CHKRbound} follows with a small change in the constant $C_0$.)

Let  $\kappa>1$
and  $L\in \N_B$ be such that $ L >2(L_B  + \kappa R)$.
If $x \in \Lambda_{L_B}(0)$, 
it follows from Lemma~\ref{lemcrucial} and \eqref{CHKRbound} that
\begin{equation}
\begin{split}
\Pi_{n,L}{\chi_{x,\eps}}\Pi_{n,L} &=
 \Phi_L\I_L^*\Pi_{n}{\chi_{x,\eps}} \Pi_{n} \I_L \Phi_L + 
\mathcal{E}_{2:x,\eps,n,L}\\
& \quad \ge C_0
  \Phi_L\I_L^*\Pi_{n} (\chi_{{x,R}} -
\eta\chi_{{x,\kappa R}})\Pi_{n} \I_L \Phi_L
+ 
\mathcal{E}_{2;x,\eps,n,L}\\
& \quad = C_0
  \Pi_{n,L} (\chi_{{x,R}} -
\eta\chi_{{x,\kappa R}})\Pi_{n,L} 
+ 
\mathcal{E}_{x,\eps,R,\kappa,n,L},
\end{split}
\end{equation}
and hence we have \eqref{boundlem} and  \eqref{PiL3} for
 $x \in \Lambda_{L_B}(0)$.  For arbitrary $x\in\Lambda_{L}(0)$,
we pick $a_x \in \Z^2_B$ such that $ x- a_x \in \Lambda_{L_B}(0)$
(such $a_x$ always exists).  Since 
$ \widehat{\chi}_{{x,\ell}}=
\widehat{U}_{a_x} \widehat{\chi}_{{x-a_x,\ell}} \widehat{U}_{a_x}^*$
 for $\ell <L$ and 
$\widehat{U}_{a_x} \Pi_{n,L} \widehat{U}_{a_x}^*=\Pi_{n,L}$,
\eqref{boundlem} and and  \eqref{PiL3} follows with   $\Ec_{n,x,L}= 
\widehat{U}_{a_x}\Ec_{n,x-a_x,L}\widehat{U}_{a_x}^*$.
\end{proof}

We can now finish the proof of of Theorem~\ref{thmWegner}.
Let 
\begin{equation}\label{tildeVL}
\widetilde{V}_{L}(x):=  \sum_{i\in \widetilde{\Lambda}_{L-\delta_u}(0)}
u(x-i)\ge
u^- \sum_{i\in\widetilde{\Lambda}_{L-\delta_u}(0)}  \chi_{i,\eps_u}  .
\end{equation}
We fix $R > 1 + 2\delta_u $, in which case
$\sum_{i\in\widetilde{\Lambda}_{L-\delta_u}(0)}  \widehat{\chi}_{i,R}
\ge  \chi_{0,L}$, and  $\kappa >1$, and pick $\eta >0$ such that
$\eta \sum_{i\in\widetilde{\Lambda}_{L-\delta_u}(0)}
  \widehat{\chi}_{i,\kappa R} \le \frac 1 2  \chi_{0,L}$.   It follows from  \eqref{tildeVL}
and
Lemma~\ref{lemWegner} that for  all $L \in \N_B$ with
 $ L >2(L_B  + \kappa R)$ we have 
\begin{align}\notag
\Pi_{n,L}\widetilde{V}_{L}\Pi_{n,L} &\ge
u^- C_0 \sum_{i\in\widetilde{\Lambda}_{L-\delta_u}(0)}
 \Pi_{n,L} (\widehat{\chi}_{{i,R}} -
\eta\widehat{\chi}_{{i,\kappa R}})\Pi_{n,L} + \Pi_{n,L} \Ec_{n,L} \Pi_{n,L}\\
&\ge \frac   {u^- C_0} 2  \Pi_{n,L} + \Pi_{n,L} \Ec_{n,L}\Pi_{n,L}\ge 
C_1 \Pi_{n,L}  \label{CHK3.1}
\end{align}
for $L \ge L^*$ for some $L^*= L^*_{n,B,\eps,R,\kappa,\eta}< \infty$ and
 $C_1 = \frac   {u^- C_0} 4$, since the error term $\Ec_{n,L}$ satisfies
\begin{equation}
\norm{\Ec_{n,L}}  \le 2 L^2 C_{n,B,\eps,R,\eta} \,\e^{-m_{n,B} L} .
\end{equation}

Theorem~\ref{thmWegner} now follows by \cite[Proof of Theorem 1.2]{CHK},
since \eqref{CHK3.1} for all $n=1,2,\dots$ gives the crucial
estimate \cite[Eq. (3.1)]{CHK}
\end{proof}

\section{The small disorder limit}
\label{sectproof}

\begin{proof}[Proof of Corollary~\ref{corlimit2}] Note first that
$ 1 < c_{b,\lambda} \le 2$ for $\lambda \le \lambda_1$, which
 we assume from now on. Fixing $B>b$,  we have \eqref{splandau2} 
with $\I_n({B,\lambda})=\I_n({B}):=\I_n({B,1})$ for all $\lambda$ and
$n=1,2,\dots$.
 By the hypothesis
on the density $\rho$, for  all $\eps >0$ we have
\begin{equation}
\nu_\lambda(\{ \abs{u} \ge \eps\})\le 
C_1 \left({\lambda} {\eps}^{-1}\right)^{\gamma-1}.
\end{equation}
  Let $L_0 \in \N_B$ (see \eqref{LB}), and let $H_{B,\lambda,0,L_0,\omega}$
and $V_{0,L_0,\omega}$ be as in \eqref{landauh2} 
 with $\lambda=1$ but
with $\nu_{\lambda}$ being the common probability distribution of the random
variables $\{\omega_i; \, i \in \Z^2\}$.  The spectrum of these finite 
volume Hamiltonians  satisfies \eqref {splandau} (appropriately modified)
 for each $\omega$, and hence  
\begin{align}\notag
&\P\left\{\sigma(H_{B,\lambda,0,L_0,\omega})\subset 
 \bigcup_{n=1}^\infty 
 [B_n - \eps, B_n +\eps]  \right\}\ge 
\P\left\{ |\omega_i|\le \eps  \ 
\text{if ${i\in  \widetilde{\Lambda}_{L_0-\delta_u}(0)}$}\right\}\\
\quad & \ge 
\left(1 - C_1 \left({\lambda} {\eps}^{-1}\right)^{\gamma-1} 
\right)^{(L_0- \delta_u)^2} \ge
 1 -C_2 \left({\lambda} {\eps}^{-1}\right)^{\gamma-1}L_0^2 \label{nospec}
\end{align}
for small $\left({\lambda} {\eps}^{-1}\right)^{\gamma-1}$.

We now apply the finite volume criterion for localization given in
 \cite[Theorem~2.4]{GK3},  in the same way  as in
 \cite[Proof of Theorem~3.1]{GK3},  with parameters (we fix $q \in ]0,1]$) 
$\eta_{I,\lambda}=\frac 1 2 \eta_{B,\lambda,I,q} = \frac 1 2 \eta_{B,1,I,q}$ and 
$Q_{I,\lambda}= Q_{B,\lambda,I,q}\le 2 \lambda^{-1} Q_{I,1}$, 
where $\eta_{B,\lambda,I}$ and $Q_{B,\lambda,I,q}$ come from 
Theorem~\ref{thmWegner}. (Note that the fact that
we work with length scales $L \in \N_B$ instead of $L \in 6\N$ only affects
 the values of the constants in \cite[Eqs. (2.16) -(2.18)]{GK3}.) 
The SLI constant $\gamma_{I,B,\lambda}$ is uniformly bounded in
closed intervals $I$ if  $\lambda\le B$. Since we are working in spectral gaps,
we use the Combes-Thomas estimate of    \cite[Proposition 3.2]{BCH} (see also
\cite[Theorem 3.5]{KK1}--its proof, based on \cite[Lemma 3.1]{BCH}, 
also works for Schr\"odinger operators with magnetic fields),
adapted to finite volume as in  \cite[Section~3]{GK3}. 

Now fix $n \in \N$, take $I= \I_n({B})$, and set $L_0=L_0(n,B)$ to
 be the smallest 
$L \in \N_B$ satisfying \cite[Eq. (2.16)]{GK3}. Let
 $E \in \mathcal{I}_n(B), \, \abs{E-B_n} \ge 2 \eps$,
 where $\eps= \eps(n,B,\lambda)) >0$ will be chosen later. Then, using
\eqref{nospec} and the Combes-Thomas estimate, we conclude that  
condition \cite[Eq. (2.17)]{GK3} will be satisfied at energy $E$
 if 
\begin{gather}
 {\eps}\ge C_3\,  {\lambda} L_0^{\frac 2 {\gamma -1}},\\
 C_4 \left(\lambda \eps\right)^{-1}L_0^{\frac {25} 3} 
\e^{- C_5 \sqrt{\eps}L_0} < 1,
\end{gather}  
for appropriate constants $C_j=C_j(n,B)$, $j=3,4,5$, with $C_5 > 0$.
This can be done by choosing
\begin{equation}
\eps = C_6 \lambda^{\frac {\gamma -1} \gamma}
 \abs{\log \lambda}^{\frac 2 \gamma},
\end{equation}
with a sufficiently large constant  $C_6=C_6(n,B)$ and taking 
$\lambda \le \lambda_2$ for some $0 < \lambda_2=\lambda(n,B, C_6)$.
We conclude from \cite[Theorem~2.4]{GK3} that
\begin{equation}\label{smalldisorder}
\left\{  E \in \I_n({B}); \  \abs{E-B_n} \ge 2C_5 \lambda^{\frac {\gamma -1} \gamma}
 \abs{\log \lambda}^{\frac 2 \gamma}\right\}\subset  
\Xi_{B,\lambda}^{\text{DL}}.
\end{equation}
for all $\lambda \le \lambda_2$.  

The existence at small disorder of dynamical mobility edges
 $\widetilde{E}_{j,n}(B,\lambda)$, $j=1,2$, 
satisfying   \eqref{loctildeE}, \eqref{abc},
and \eqref{lambda20}  now follows from Theorem~\ref{thmdeloc}
and \eqref{smalldisorder}.

The case when $\e^{\abs{u}^\alpha} \rho(u)$
is bounded for some  $\alpha>0$ can be treated in a similar way.
\end{proof}

\end{document}